\documentclass[twocolumn,showpacs,superscriptaddress,preprintnumbers,amsmath,amssymb]{revtex4}
\usepackage{graphicx}
\usepackage{dcolumn}
\usepackage{natbib}
\usepackage{color}
\usepackage{bm}
\def\be{\begin{equation}}
\def\ee{\end{equation}}
\def\bea{\begin{eqnarray}}
\def\eea{\end{eqnarray}}
\def\f{\frac}

\def\nn{\nonumber}
\def\ra{\rangle}
\def\la{\langle}

\def\tB{{\widetilde{B}}}

\begin{document}

\title{Quantum time of arrival distribution in a simple lattice model }

\author{Shrabanti Dhar}
\affiliation{University of Calcutta,
92 Acharya Prafulla Chandra Road, Kolkata 700009, India}
\author{Subinay Dasgupta}
\affiliation{University of Calcutta,
92 Acharya Prafulla Chandra Road, Kolkata 700009, India}
\author{Abhishek Dhar}
\affiliation{International Centre for Theoretical Sciences, TIFR, Bangalore 560012, India}

\begin{abstract}
Imagine an  experiment where a quantum particle inside a box is released at some time  in some initial state. A detector is placed at a fixed location inside the box and its clicking signifies arrival of the particle at the detector. What is the \emph{time of arrival} (TOA) of the particle at the detector ?  
Within the paradigm of the measurement postulate of quantum mechanics,
one can use the idea of projective measurements to define the TOA.
We consider the setup where a detector keeps making instantaneous 
measurements at regular finite time intervals {\emph{till}} it detects the particle at some time $t$, which is defined as the TOA. This is a stochastic variable and, 
for a simple lattice model of a  free particle in a one-dimensional box, we find interesting features such as power-law tails  in its distribution and in the probability of survival (non-detection). We propose a perturbative calculational approach which yields results that compare very well with  exact numerics.
\end{abstract}


\maketitle

The problem of defining the  time of arrival (TOA) of a particle in quantum mechanics, and determining its probability distribution, has been a difficult and 
intriguing problem, one  that is closely related to the foundations of quantum mechanics. 
A  large body of work has studied this problem using a wide variety of approaches  \cite{allcock69,kijowski74,kumar85,grott96,aharonov98,savvidou06,mugap00,damborenea02,galapon04,galapon05,muga08,yearsley,savvidou12,vona13}. A somewhat older but still relevant review of the various attempts to do so have been described  in \cite{muga00}. 
Experimentally, time of flight of atoms from source to detector, are routinely measured but these are typically  in the semi-classical regime, and making meaning of  these measurements in the  quantum regime is not straightforward \cite{muga00}.

 There are several aspects that are involved  in discussions of the TOA :\\
{(i)} First there is  the question of the effect of measurements made to detect the particle's arrival. The question of repeated ideal measurements of a quantum system was discussed in the seminal paper of Misra and Sudarshan \cite{misra77} who studied this question in a general setting  and showed the surprising result, the so-called quantum Zeno effect, that the probability of detecting a particle (or \emph{decay} from the initial state) vanishes in the limit  that the time interval between measurements $\tau \to 0$ \cite{shimizub05,facchi08}.  This means that continuous measurements to find the time of arrival leads to the particle  \emph{being never detected}! The Zeno effect has been experimentally verified \cite{wineland90} though questions of interpretation remain \cite{itano09}. Hence the question of making measurements at regular finite intervals arises  and it becomes  necessary 
to study the effect, that null measurements have, on the time evolution of a quantum system and on the TOA distribution \cite{savvidou06}.  A related issue is that of  defining POVMs corresponding to TOA measurements \cite{vona13};\\ 
{(ii)} There is then the question of defining a self-adjoint time operator and some progress has been made here \cite{galapon04}.  Determining arrival time distributions from these definitions has its own issues~\cite{galapon05}; \\
{(iii)} Finally there is the important question of trying to connect to real  experiments. One then needs to incorporate into the picture the entire 
measurement process by also modeling the measuring device and its interaction with the particle. This has been discussed in, for example, \cite{mugap00,damborenea02,savvidou12}.

\begin{figure}
\includegraphics[clip,width=8cm, angle=0]{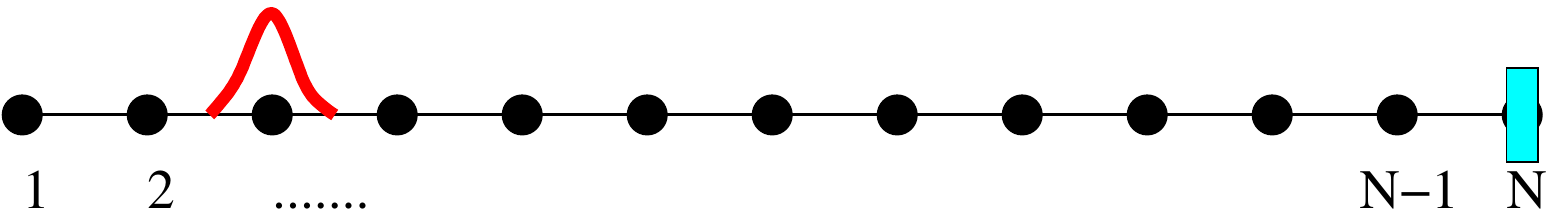}
\caption{ A schematic  representation of the simplified $1D$ lattice model studied in this paper.  A particle inside a  box  is released in a fixed initial wavefunction and a detector at site $N$ makes instantaneous measurements  to see if the particle has arrived at site $N$. The experiment ends once the particle is detected and the time of arrival (TOA) is recorded.}
\label{schematic}
\end{figure}

In this Letter, our focus is on aspect (i), namely we discuss the distribution of time of arrivals resulting from repeated ideal measurements made at regular \emph{finite} time intervals.  In particular,  with the aim of  being able to explicitly compute the TOA distribution, we study a  lattice version of a free 
particle in one-dimension.
We consider a quantum particle that is  prepared in a given 
initial  state at some time instant (say $t=0$) and  a detector is placed at some fixed location [schematically shown in Fig.~(\ref{schematic})]. 
 The detector makes instantaneous quantum
measurements at regular intervals of time $\tau$, and keeps doing so {\emph{till it detects the particle}}, say on the $n^{\rm th}$ observation,  at time $t=n\tau$. This is taken to be the TOA, which is a stochastic variable. The time evolution of the system undergoing repeated measurements constitutes a non-unitary dynamics.  
Here we examine the survival probability   $P(t)$ that the particle is un-detected till time $t$. The limit of continuous measurements $\tau\to 0$ gives $P(t) \to 1$ but we will see that any finite $\tau$ leads to a non-trivial survival probability with interesting  power-law tails.

We note that  a closely related work is that by Anastopoulos and Savvidou \cite{savvidou06} who consider a free particle  on the infinite real line. The particle is initially prepared in the negative half line and then subjected to regular measurements, that correspond to projections onto the positive half space. The approach followed in this Letter is similar to their paper, however, while their main emphasis was in trying to understand the $\tau \to 0$ limit and the related problems, here we focus on the \emph{finite} $\tau$ problem.  
Our study is  also different from earlier studies on the effect of  finite-time-interval measurements  on the Zeno effect which have typically looked at few-level unstable systems and examined their decay and survival probabilities \cite{shimizub05,facchi08,hegerfeldt96}. In contrast, our set-up is that of an extended system, where measurements are made in part of the space and the system's time evolution is altered by the measurements.  
Another related study is that of Yi \emph{et al} \cite{ingold11} who consider the effect of 
multiple measurements involving observation at a single site,  on the density matrix of a particle in a one-dimensional box. However, there the focus was not on first arrival, and the effective evolution equation of the density matrix is completely different from that considered by us.

 Our model consists of  a particle moving on a discrete lattice of $N$ sites 
and its dynamics is described  by a tight-binding type Hamiltonian of the form
\bea
{H}= \sum_{\substack{\ell,m =1 }}^N H_{\ell,m}~|\ell\ra \la m|
\eea
where $H$ is a symmetric matrix. The free time evolution of  $|\psi\ra $ is given by
$|\psi (t) \ra = U^t |\psi (0)\ra,~~{\rm where}~~U^t=e^{-i H t/\hbar}~.$
Let us define the projection operator $ {A}= \sum_{j \in D} |j \ra \la j|$ corresponding to a measurement to detect the particle in the domain $D$ containing a fixed set of sites, and the complementary operator $B=1-A$. According to the  measurement postulate of quantum mechanics, the probability of detecting the particle on performing a measurement on the state $|\psi \ra$  is $ p =  \sum_{j \in D} | \la j | \psi \ra |^2 = \la \psi | A | \psi \ra$. The probability of non-detection or the {\emph{ survival probability}} is then $P= \la \psi | B | \psi \ra  =1-p$.
The measurement postulate also tells us that measurements  alter the Hamiltonian time evolution of the system. 
Thus if a measurement  {\emph{does not}} detect the particle, then the wavefunction immediately after a measurement projects to $|\psi^+\ra= B|\psi\ra/\sqrt{P}$.

We now consider a sequence of measurements $n=1,2\ldots$ at intervals of time $\tau$ which continue until a particle is detected. \emph{Thus the time evolution is given by a sequence of unitary evolutions followed by projections into the subspace corresponding to $B$ till the particle is detected}.    
Let $|\psi^-_n\ra$ and $|\psi^+_n\ra$ be the wave functions (un-normalized) of the system, immediately before and after  the $n^{\rm th}$ measurement respectively. We note that $ 
|\psi^-_n\ra = U^\tau|\psi^+_{n-1}\ra$ and $|\psi^+_n\ra=B|\psi^-_n\ra$. Hence, defining $\tB=BU^\tau$, it follows that
\bea
|\psi^-_n \ra =   U^\tau \tB^{n-1} |\psi (0) \ra~, ~~|\psi^+_n \ra =   \tB^{n} |\psi (0) \ra~. \label{evolpsi}
\eea
Let $P_n$ be the probability of survival after  $n$ measurements.
Then clearly
\bea
P_1 =  \la \psi^-_1|B | \psi^-_1 \ra= \la \psi(0)| \tB^\dagger \tB |\psi(0) \ra=\la \psi^+_1|\psi^+_1\ra~. \nn
\eea
Note that $P_1$ is thus the normalizing factor for $|\psi^+_1\ra$ and also for $|\psi^-_2\ra$.  
The survival probability after the second measurement is obtained as the product of non-detection at $n=1$ times the probability of non-detection at $n=2$ and this is
\bea
P_2 =P_1 \times \f{ \la\psi^-_2 |}{\sqrt{P_1}} B \f{|\psi^-_2 \ra}{\sqrt{P_1}}
=\la \psi(0)| \tB^{\dagger 2} \tB^2 |\psi(0) \ra=\la \psi^+_2|\psi^+_2\ra~.\nn
\eea
Proceeding iteratively in this way, we get 
\bea
P_n =\la \psi(0)| \tB^{\dagger n} \tB^n |\psi(0) \ra=\la \psi^+_n|\psi^+_n\ra~.\label{PS}
\eea
If we  imagine an ensemble of identically prepared states, on which we perform repeated measurements, then $P_n=\la \psi^+_n|\psi^+_n \ra$ gives the fraction of systems for which there has been no detection and that are still evolving.
Note that the difference $P_{n-1}-P_n$ gives the probability of first detection in the $n^{\rm th}$ measurement.

 In the rest of the paper, we shall  consider the special case of a one-dimensional lattice where the measurement is made at a single site $N$ and the corresponding projection operator is thus ${A}=|N\ra \la N|$.  
In the position basis, the complementary operator $B$  corresponds to an $N \times N$ matrix  with elements ${B}_{jk}=\delta_{j,k} ~(1-\delta_{j,N} )$.
Our main interest will be in the survival probability, $P_n$  (or equivalently, $P(t)$, where $t=n\tau$) given by Eq.~(\ref{PS}).
An explicit solution of this problem requires one to diagonalize the non-Hermitian evolution operator $\tB$ which in general is difficult. 

 We study a Hamiltonian that incorporates nearest neighbor hopping of a 
particle and, first consider the case of an open chain, corresponding to a free particle inside a $1D$ box. Thus, the Hamiltonian is given by
\bea 
H = - \sum_{k=1}^{N-1} \gamma \left( ~|k+1 \rangle \langle k | +  |k \rangle \langle k +1 | ~\right)~. 
\eea
Without loss of generality we can set $\gamma=1, \hbar=1$. 
The eigenvalues and eigenvectors of this Hamiltonian are given by
$\epsilon_s=-2\cos [s\pi/(N+1)]$ and  $\psi_s(\ell)=[2/(N+1)]^{1/2} \sin [ s \ell \pi/ (N+1)]$ with $s=1,2,\ldots,N$. 
The  orthogonal matrix $V$ with matrix elements $V_{\ell s}=\psi_s(\ell)$ diagonalizes $H$ and the time-evolution of the state, from $|\psi^+_{n-1}\ra$ to    $|\psi^-_{n}\ra=U^\tau |\psi^+_{n-1}\ra $ with $U_{\ell,m}^\tau=\sum_{s} V_{\ell,s} e^{-i \epsilon_s \tau} V_{m,s}$, is easy to implement numerically.
The projection to $|\psi^+_n \ra = B|\psi^-_{n}\ra$  is then simply obtained, in the position basis, as 
\bea
\psi^+_{n}(\ell) = \left\{ \begin{array}{ll}
                     \psi^-_{n}(\ell) &\mbox{~~~{\rm for} ~~$\ell \neq N~$,} \\
                                   0  &\mbox{~~~{\rm{for}~~ $\ell=N$}~}\end{array} \right.\label{am} 
\eea
Thus, numerically it is easy to start with any initial wavefunction $\psi^+_0(\ell)$ and evolve it using the above iteration scheme.

We now present a perturbative calculation of the time-evolution of the wavefunction and of the survival probability. The small parameter here is the time $\tau$ between successive measurements (compared to the time for the wavefunction to spread which is $\hbar/\gamma$).  
Let us use the notation $H_N$ to denote the  Hamiltonian matrix on a $N$-site lattice and let $h_{N}=(0,0,\ldots,1)^T$ be a $N$ dimensional column vector with only the last element non-zero. We note that the vector $e^{i H \tau} h_N$ is an exact eigenstate of $\tB$ with eigenvalue $0$  and find the other eigenvalues and eigenstates perturbatively. Expanding $\tB$ to second order in $\tau$ we have
\begin{align}
\tB &= B~(I-iH \tau-H^2 \tau^2/2+\ldots )~\nn \\
&=
\left( \begin{array}{cc}
I_{N-1} & 0 \\
0 & 0 
\end{array} \right)~
\left[ \left( \begin{array}{cc}
I_{N-1} & 0 \\
0 & 1 
\end{array} \right)-i \tau
\left( \begin{array}{cc}
H_{N-1} & -h_{N-1} \\
-h_{N-1}^T & 0 
\end{array} \right)~ \nn \right. \\  
&- \left. \f{\tau^2}{2}
\left( \begin{array}{cc}
H^2_{N-1}+h_{N-1} h^T_{N-1} & -H_{N-1} h_{N-1} \\
-h_{N-1}^T H_{N-1}  & 1 
\end{array} \right)~\right] \nn \\
&= \left( \begin{array}{cc}
I_{N-1} -i \tau H_{N-1} - \tau^2 H^2_{N-1}/2 + Z_{N-1} & C_{N-1} \\
0 & 0 
\end{array} \right)~, \nn 
\end{align}
where $I_N$ denotes a $N$-dimensional unit matrix, $Z_{N-1}= -(\tau^2/2) h_{N-1} h^T_{N-1}= -(\tau^2/2)|N-1\ra \la N-1|$ is a $(N-1) \times (N-1)$ matrix with one non-vanishing element and $C_{N-1}= i\tau h_{N-1}+(\tau^2/2) H_{N-1}h_{N-1}$ . 
Let the $N-1$ eigenstates and eigenvalues of the matrix $Q_{N-1} = I_{N-1} -i \tau H_{N-1} - \tau^2 H^2_{N-1}/2 + Z_{N-1}$ be denoted by $|\chi_s \ra$ and $\mu_s$ respectively, satisfying
\bea
Q_{N-1} | \chi_s \ra =\mu_s |\chi_s \ra~.
\eea
Denoting the components $\chi_s(\ell)=\la \ell |\chi_s\ra$, it is easily seen that the vectors $(\chi_s(1),\chi_s(2),\ldots,\chi_s(N-1),0)$ form the remaining $(N-1)$ eigenstates of $\tB$. We now find $\chi_s(\ell)$ and $\mu_s$ using perturbation theory.

The eigenfunctions and eigenvalues of $H_{N-1}$ are respectively given by $\phi_s(\ell)=[2/N]^{1/2} \sin [ s \ell \pi/ N]$ and $e_s=-2\cos(s\pi/N)$, with $s=1,2,\ldots,N-1$ and $\ell=1,2,\ldots,N-1$. Treating the part $Z_{N-1}$ of $Q_{N-1}$ as a perturbation, we get from  first order perturbation theory
\begin{align}
|\chi_s\ra &=|\phi_s \ra-\f{\tau^2}{2} \sum_{s'\neq s} \f{\phi_s(N-1) \phi_{s'}(N-1)}
{e_s-e_{s'}} |\phi_{s'}\ra+O(\tau^3)~, \nn \\
\mu_s &=1-i \tau e_s-\f{\tau^2}{2}e_s^2  + \la\phi_s|Z_{N-1}|\phi_s\ra+O(\tau^3)\nn \\&=e^{-i\tau e_s} e^{-\alpha_s\f{\tau}{2}}+O(\tau^3)~
\end{align}
where $\alpha_s= -(2/\tau)\la\phi_s|Z_{N-1}|\phi_s\ra=\tau\phi^2_s(N-1)=\tau \phi^2_s(1)$.
Now we can use Eq.~(\ref{evolpsi}) to find the state of the system at time $t=n \tau$, after $n$ measurements. If the initial state is an eigenstate of $H_{N-1}$, {\emph{i.e}} $|\psi_0^+\ra=|\phi_s\ra$, then we have approximately
\bea
|\psi_t^+\ra = \mu_s^n |\phi_s \ra= e^{-i t e_s} e^{-\f{\tau t}{2} \phi^2_s(1)} |\phi_s\ra= P^{1/2}_s(t) e^{-i t e_s}  |\phi_s\ra, \nn
\eea
where $P_s(t)$, the survival probability (of the $s^{\rm th}$ energy eigenstate) given by 
\begin{equation}
P_s(t)=\la \psi^+_t|\psi^+_t \ra = e^{-\alpha_s t}~.
\nn
\end{equation}
Thus, $\alpha_s$ represents the decay rate of the state $|\phi_s\ra$. We see that when the initial state is an eigenstate of the Hamiltonian, the survival probability decays exponentially with time, with the rate of decay depending on the measurement interval $\tau$ and the probability density of the wavefunction near the detection point. In the limit $\tau \to 0$, the decay rate vanishes \emph{implying the Zeno effect}.

\begin{figure}
\centering
\includegraphics[clip,width=3.2in, angle=0]{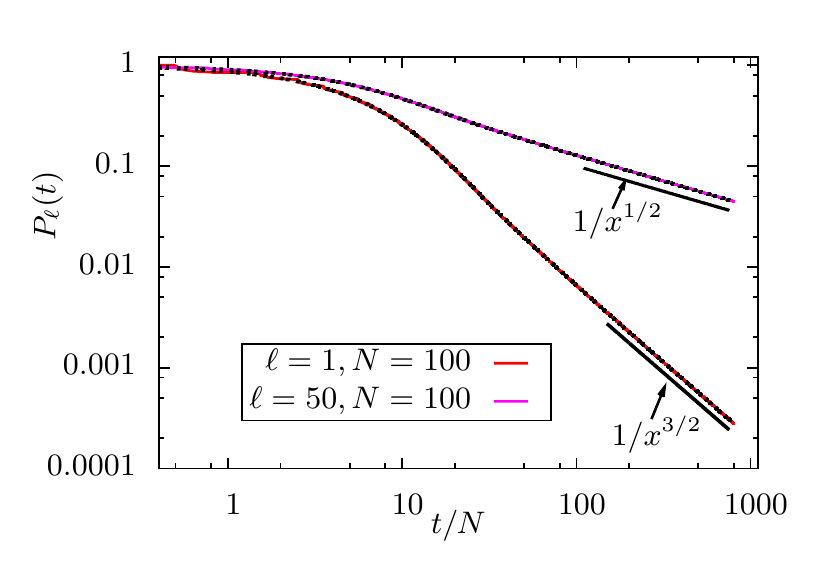}
\caption{Decay of the survival probability $P_\ell(t)$  for different initial positional eigenstates, obtained from exact numerical evaluation of the dynamics.  The black dotted lines are the predictions from perturbation theory. The solid black lines are the predicted power-law decays for bulk and boundary initial points.
The  measurement was done at site $N$  and  measurement time interval  was taken to be  $\tau=0.1$.}
\label{open}
\end{figure}

For the case where the particle is initially at site $\ell$, the  initial  position eigenstate can be expanded as 
$|\ell \ra = \sum_s c_s | \phi_s \ra$, with $c_s=\phi_s(\ell)$,  so that at time $t$ we now get
\bea
| \psi^+_t \ra = \sum_s \phi_s(\ell) P^{1/2}_s(t) e^{-i e_s t} |\phi_s \ra~.
\eea
The survival probability is then obtained as
\begin{align}
P_\ell(t)&=\la \psi_t^+ | \psi_t^+\ra = \sum_s \phi_s^2(\ell) P_s(t) \nn \\
&=\sum_{s=1}^N \f{2}{N} \sin^2 \left(\f{s \pi \ell}{N}\right) e^{-\f{2 \tau t}{N} \sin^2 (\f{s \pi}{N})}~. \label{Pit} 
\end{align}
 The difference between the $(s+1)$-th and $s$-th term of this summation will be small 
for  large $N$ and small $t\tau/N^3$. We can then convert the sum to an integral and get
\bea
P_\ell(t)  = \f{2}{\pi}  \int_0^\pi dq \sin^2 (q \ell) e^{-\f{2 t \tau}{N} \sin^2 q}~ \label{Pitcont1} 
\eea
Thus we find that $P_\ell(t)$ has the scaling form $P_\ell(t)=f(t\tau/N)$. Defining,
\bea
x \equiv \frac{t\tau}{N} ~ \label{defx}
\eea
we also see that when $x$ is large, only small values of $q$ in this integral will matter, and hence
\bea
P_\ell(t)=  \f{1}{\pi} \int_{-\infty}^{\infty} dq \sin^2 (q \ell) e^{-2xq^2} =\f{1}{\sqrt{8\pi x}}\left[1-e^{-\ell^2/2x}\right]. \nn 
\eea 
For points close to the boundary, $\ell \sim O(1)$, we get $P_\ell(t)\sim 1/t^{1/2}$ at times  $1<< t\tau/N << \ell^2$, and then a cross-over to $P_\ell(t) \sim 1/t^{3/2}$ at large times $t \tau /N >> \ell^2$.
For points in the bulk of the sample, $\ell \sim O(N)$,
 $\ell^2/x (\sim N^3/t\tau )$ is large in the time domain where $t\tau/N^3 <<1$,  hence one observes only the behaviour $P_\ell(t)\sim 1/t^{1/2}$. The power-law decay with time changes to an exponential decay at times $t \sim N^3$, when the sum in Eq.~(\ref{Pit}) is dominated by one term, namely the one corresponding to the smallest eigenvalue.  
The first arrival probability is obtained as $p(t)=-dP/dt$.
In Fig.~(\ref{open}) we show the comparison between the analytic predictions for the survival probability with the  exact numerical results.
The agreement is very good.

 We shall now discuss the case of periodic boundary  condition.  
The Hamiltonian is now given by,
\be {H_P} = -\sum_{k=1}^{N-1}\left(~ |k+1 \ra \la k | + |k \ra 
\la k+1 |~ \right)  - |1 \ra \la N |  -  | N \ra \la 1 | \ee 
For {\em even} values of $N$, there are $(N-2)/2$ eigenvalues, 
\[ \epsilon_s=-2\cos(2 s \pi/N) \]
each with two degenerate eigenvectors
\bea 
\psi_s(\ell)&  =& (2/N)^{1/2} \sin (2 s \ell \pi/ N) \nonumber \\
 \psi_{s+N/2-1}(\ell)&=&(2/N)^{1/2} \cos (2 s \ell \pi/ N) \nonumber \eea
for $s=1,2,\ldots,N/2-1$. The remaining two eigenvectors (non-degenerate) are $\psi_{N-1}(\ell)=(-1)^\ell/N^{1/2}$ and $ \psi_N(\ell)=1/N^{1/2}$ with eigenvalues $-2$ and $2$ respectively. 
However, in this case we notice that, the eigenstates $\phi_{2s}(\ell)$ of $H_{N-1}$~(with open boundary condition) for $s=1,2,\ldots,N/2-1$ are also exact eigenstates of 
$H_P$ and they all vanish at the site $N$. Hence these are also exact eigenstates of $\tB(=Be^{-i\tau H_P)}$ with eigenvalue $e^{-i e_{2s} \tau}$ and \emph{ do not decay}. Also, one observes that the vector $e^{i\tau H_P}\left(0,0,....1\right)^{T}$ is an exact eigenvector of $\tB$ with eigenvalue zero; however, this eigenvector does not contribute to dynamics as the eigenvalue is only {\em zero}. Thus, we now have $N/2$ exact eigenvectors of $\tB$. The remaining eigenvectors can be found perturbatively as before.  
We note that writing Hamiltonian $H_P$ in block form now gives us the same form as $H_{N-1}$ before, 
while the vector $h_{N-1}$ now has the form $(1,0,\ldots,0,1)^{T}$ and $Z_{N-1}$ is therefore given by
\vspace{-0.2cm}
\bea Z_{N-1}=-(\tau^2/2)h_{N-1}h_{N-1}^T =-(\tau^2/2)(|1\rangle\langle1|+ ~ \nonumber \\
\hspace{-1.0cm} |1\rangle\langle N-1|+|N-1\rangle\langle1|+|N-1\rangle\langle N-1|)~\nn\eea 
Let the initial state of the system be any one of $|\phi_{2s+1}\ra$, $s=0,1,\ldots N/2-1$.
The decay rate is then given by
\[ \alpha_{2s+1} =-(2/\tau)~\la\phi_{2s+1}|Z_{N-1}|\phi_{2s+1}\ra
=4 \tau \phi^2_{2s+1} (1).\] 
Thus, initial eigenstates which are symmetric about  the centre ($N/2$) of the ring decay with this rate, while the odd states remain undetected and do not decay. 

If the initial state is a position eigenstate ($\ell \neq N$) then we  expand in the basis of $H_{N-1}$ and obtain
\begin{align}
| \psi^+_t \ra &= \sum_{s=1}^{N/2-1} \phi_{2s}(\ell) e^{-i e_{2s} t} |\phi_{2s} \ra \nn \\
& + \sum_{s=0}^{N/2-1} \phi_{2s+1}(\ell) P^{1/2}_{2s+1}(t) e^{-i e_{2s+1} t} |\phi_{2s+1} \ra~. \nn
\end{align}
Taking the inner product  we get
$P_\ell(t)-P_\ell(\infty)=\sum_{s=0}^{N/2-1} \phi_{2s+1}^2(\ell) P_{2s+1}(t)~$
where $P_\ell(\infty)= \sum_{s=1}^{N/2-1} \phi_{2s}^2(\ell)$ is 
$1/2$  for $\ell \neq N/2$ and vanishes for $\ell = N/2$.
Thus,  for all initial position eigenstates except $\ell=N/2$,  the survival probability is $P_\ell(\infty)=1/2$. 
As for the case of an open chain, for large $N$ and small $t\tau/N^3$ we can 
convert the sum to an integral to get
\bea P_\ell(t)-P_\ell(\infty) = \f{1}{\pi} \int_0^\pi dq \sin^2 (q \ell) 
e^{-( 8 t \tau/N) \sin^2 q}~. \label{Pitcont1} \eea 
For large $t\tau/N(=x,$ say), this integral becomes
\bea P_\ell(t)-P_{\ell}({\infty})  
=\f{1}{ 8 \sqrt{ 2\pi x}}\left[1-e^{-\ell^2/ 8x}
\right]. \label{Pitcont2} \eea

Thus, as before we find that, for initial starting points close to detector  [$\ell \sim O(1)$ or $N-\ell \sim O(1)$],  $P_\ell(t)$ decays  to its asymptotic limiting value as $\sim 1/t^{3/2}$, while for  initial starting points in the bulk of the sample we get a decay as $\sim 1/t^{1/2}$.  It can also be shown that  for $\ell=N$ we get $P_N(\infty)=0$ and a $1/t^{1/2}$ decay. 
We note here that a  recent paper \cite{mallick13}  considered the motion of a quantum particle on a ring in the presence of trapping sites (modeled by non-Hermitian potentials) and find similar results.

{\bf Conclusion:}  In this Letter we considered the example of a particle
inside a box, that is released at time $t=0$ with an initial wave
function $|\psi(0)\ra$, which could either be an extended energy
eigenstate or a spatially localized state. A detector placed at a
fixed location is turned on  at regular small intervals of time
$\tau$ and makes instantaneous measurements. The first click of the detector, say on the $n$-th
measurement, gives the  time of arrival (or time of first detection) $t=n\tau$.  One can imagine
an ensemble of such experiments being performed, such that once a
particle is detected in any of the realizations, it is not studied
anymore and we carry on with the remaining realizations. Thus each
different realizations of the experiment ends at a different time and
we get a distribution of times. For this process, the probability
distribution of the time of arrival and the corresponding survival probability,
can be defined un-ambiguously according to the measurement postulates of
quantum mechanics.  The effective time evolution constitutes an
interesting example of non-unitary time evolution for which we show
that an accurate solution is given from standard perturbation
theory. Using this, we obtained non-trivial results for the survival
probability for a simple lattice Hamiltonian model of a free particle.
Interesting features, including non-trivial power-law tails of the
survival probability, are observed and these feature are different
from the first passage behaviour in a classical system \cite{redner}.

Our formalism and results are easily extendable to more realistic systems, {\emph{e.g} those in higher dimensions with extended detectors. Cold atoms on optical  lattices would be ideal experimental systems where some of our predictions can be tested. These tests are interesting since they offer a direct test of the measurement postulate of quantum mechanics. \\

\begin{center}
{\bf ACKNOWLEDGEMENT}\\
\end{center}
One author (S. Dhar) gratefully acknowledges CSIR, India for providing the research fellowship through sanction no. 09/028(0839)/2011-EMR-I.  The work of S. Dasgupta is supported by UGC-UPE (University of Calcutta). S. Dasgupta is also grateful to ICTS, Bangalore for hospitality. AD thanks DST for support through the Swarnajayanti grant.


\begin{thebibliography}{99}

\bibitem{allcock69} G. R. Allcock, Ann. of Phys. {\bf 53}, 253 (1969); G. R. Allcock, Ann. of Phys. {\bf 53}, 286 (1969). 

\bibitem{kijowski74} J. Kijowski, Rep. Math. Phys. {\bf 6}, 361 (1974).

\bibitem{kumar85} N. Kumar,  Pramana-J. Phys {\bf 25}, 363 (1985). 

\bibitem{grott96} N. Grot, C. Rovelli, R.S. Tate,  Phys. Rev. A {\bf 54}, 4676 (1996). 

\bibitem{aharonov98} Y. Aharonov, J. Oppenheim, S. Popescu, B. Reznik, and W.G. Unruh, Phys. Rev. A {\bf 57}, 4130 (1998). 


\bibitem{savvidou06} C. Anastopoulos and N. Savvidou, J. Math. Phys. {\bf 47}, 122106 (2006). 


\bibitem{mugap00} J. G. Muga, A. D. Baute,  J. A. Damborenea, and I. L. Egusquiza, arxiv:quant-ph/0009111 (2000).

\bibitem{damborenea02} J. A. Damborenea, I. L. Egusquiza, G.C. Hegerfeldt, and J.G. Muga,  Phys. Rev. A {\bf 66}, 052104 (2002).


\bibitem{galapon04} E. A. Galapon, R. F. Caballar, and R. T. Bahague Jr, Phys. 
Rev. Lett. {\bf 93}, 180406 (2004).

\bibitem{galapon05} E. A. Galapon, F. Delgado, J. G. Muga, I. Egusquiza, Phys. Rev. A {\bf 72}, 042107 (2005). 




\bibitem{muga08}  J. Echanobe, A. del Campo, J. G. Muga, Phys. Rev. A {\bf 77}, 032112 (2008). 

\bibitem{yearsley} J. M. Yearsley, D. A. Downs, J. J. Halliwell, and A. K. Hashagen, Phys. Rev. A {\bf 84}, 022109 (2011).



\bibitem{savvidou12} C. Anastopoulos and N. Savvidou, Phys. Rev. A {\bf 86}, 012111 (2012).

\bibitem{vona13} N. Vona, G. Hinrichs and D. D\"{u}rr, Phys. Rev. Lett. {\bf 111}, 220404 (2013).


\bibitem{muga00} J.G Muga and C.R. Leavens, Phys. Rep. {\bf 338}, 353 (2000).



\bibitem{misra77} B. Misra, E.C.G. Sudarshan, J. Math. Phys. {\bf 18}, 756 (1977).

\bibitem{shimizub05} K. Koshino and A. Shimizu, Phys. Rep. {\bf 412}, 191 (2005). 

\bibitem{facchi08} P. Facchi and S. Pascazio, J. Phys. A {\bf 41}, 493001 (2008).



\bibitem{wineland90} W. M. Itano, D. J. Heinzen, J. J. Bollinger, and D. J. Wineland, Phy. Rev. A {\bf 41}, 2295 (1990); A. G. Kofman and G. Kurizki, Phys. Rev. A {\bf 54}, R3750 (1996); P.G. Kwiat \emph{et al.}, Phys. Rev. Lett {\bf 83}, 4725 (1999); J. I. Cirac, A. Schenzle, and P. Zoller, Euro. Phys. Lett. {\bf 27}, 123 (1994).

\bibitem{hegerfeldt96}  G. C. Hegerfeldt and D. G. Sondermann, Quant. Semiclassic. Opt. {\bf 8}, 121 (1996).

\bibitem{itano09} W.M. Itano, J. Phys. : Conf. Series, {\bf 196}, 012018 (2009).
  
\bibitem{ingold11} J. Yi, P. Talkner, and G.-L. Ingold, Phys. Rev. A {\bf 84}, 032121 (2011).

\bibitem{mallick13} P.L. Krapivsky, J.M. Luck and K. Mallick,  J. Stat. Phys. {\bf 154}, 1430(2014). 

\bibitem{redner} S. Redner, {\em A Guide to First Passage Processes} (Cambridge University Press, Cambridge, 2001) page 40ff.


\end{thebibliography}
\end{document}